\documentclass[runningheads]{llncs}
\usepackage{cite}
\usepackage{todonotes}
\usepackage[utf8x]{inputenc}

\usepackage{algorithm}
\usepackage{algpseudocode}
\usepackage{listings}
\usepackage{enumitem}

\usepackage{endnotes}
\let\footnote=\endnote

\makeatletter
\renewcommand\section{\@startsection{section}{1}{\z@}%
                       {-6\p@ \@plus -7\p@ \@minus -7\p@}%
                       {6\p@ \@plus 7\p@ \@minus 7\p@}%
                       {\normalfont\large\bfseries\boldmath
                        \rightskip=\z@ \@plus 7em\pretolerance=10000 }}
\renewcommand\subsection{\@startsection{subsection}{2}{\z@}%
                       {-7\p@ \@plus -6\p@ \@minus -6\p@}%
                       {7\p@ \@plus 6\p@ \@minus 6\p@}%
                       {\normalfont\normalsize\bfseries\boldmath
                        \rightskip=\z@ \@plus 7em\pretolerance=10000 }}
\renewcommand\subsubsection{\@startsection{subsubsection}{3}{\z@}%
                       {-6\p@ \@plus -8\p@ \@minus -8\p@}%
                       {-2em \@plus -0.22em \@minus -0.1em}%
                       {\normalfont\normalsize\bfseries\boldmath}}

\makeatother

\setlist[itemize]{leftmargin=*, topsep=0pt}

\newcommand\schema[1]{{\normalfont\fontfamily{cmvtt}\selectfont #1}}

\newcommand\prefix[2]{{\normalfont\fontfamily{cmvtt}\selectfont #1:\allowbreak #2}}

\begin{document}
\title{Simple-ML: Towards a Framework for \\ Semantic Data Analytics Workflows}

\author{
Simon Gottschalk\inst{1} \and
Nicolas Tempelmeier\inst{1} \and
Günter Kniesel\inst{2} \and \\
Vasileios Iosifidis\inst{1} \and
Besnik Fetahu\inst{1} \and
Elena Demidova\inst{1}
}

\institute{
L3S Research Center, Leibniz Universität Hannover, \email{\{lastname\}@L3S.de}
\and
Smart Data Analytics Group (SDA), Universität Bonn, \email{guenter.kniesel@uni-bonn.de}
}

\authorrunning{Simon Gottschalk et al.}

\maketitle

\begin{abstract}

In this paper we present the Simple-ML framework that
we develop to support efficient configuration, robustness and reusability of data analytics workflows through the adoption of semantic technologies. 
We present semantic data models that lay the foundation for the framework development and discuss the data analytics workflows based on these models. Furthermore, we present an example instantiation of the Simple-ML data models for a real-world use case in the mobility domain. 


\end{abstract}

\section{Introduction}
\label{sec:introduction}

The creation of a \textit{Data Analytics Workflow} (DAW) demands significant data science expertise. This expertise is required to 
integrate data from heterogeneous sources, to extract features for \textit{machine learning} (ML) tasks, to configure the DAW and to optimize its parameters. The Simple-ML framework, which we currently develop to address these challenges, aims to enable a robust, efficient and reusable DAW configuration through seamless integration of semantic information in all  typical DAW components, making it a \textit{Semantic Data Analytics Workflow} (SDAW). The adoption of semantic information, such as a domain model and semantic dataset profiles, substantially differentiates Simple-ML from existing data science frameworks such as 
RapidMiner or Microsoft Azure. 

In this paper we present Simple-ML and illustrate its adoption to data analytics for urban mobility. Popular problems in this domain include short-term road traffic forecasting \cite{ijcai2018-482}, the prediction of congestion patterns \cite{DBLP:journals/tbd/Nguyen0C17} and impact prediction of planned special events \cite{Tempelmeier-geoinf}. The corresponding SDAWs require a variety of heterogeneous data sources, including but not limited to traffic and mobility data streams, map data (e.g. OpenStreetMap), knowledge graphs containing events and spatial entities (e.g. EventKG \cite{GottschalkD18} and Wikidata), as well as traffic warnings, accidents, weather conditions and event calendars \cite{Tempelmeier-geoinf, ijcai2018-482}.

Our contributions are as follows: (i) We propose the Simple-ML framework for SDAWs: a semantic-driven approach that aims at increasing the efficiency of the workflow configuration, as well as robustness and reusability of DAWs using semantic technologies. (ii) We introduce a domain-specific semantic data model that provides semantic descriptions of the application domain and domain-specific relevant datasets (i.e. dataset profiles). (iii) We illustrate an application of the Simple-ML framework to a real-world use case in the mobility domain.

\begin{figure}[ht]
 \centering
 \includegraphics[width=0.75\textwidth]{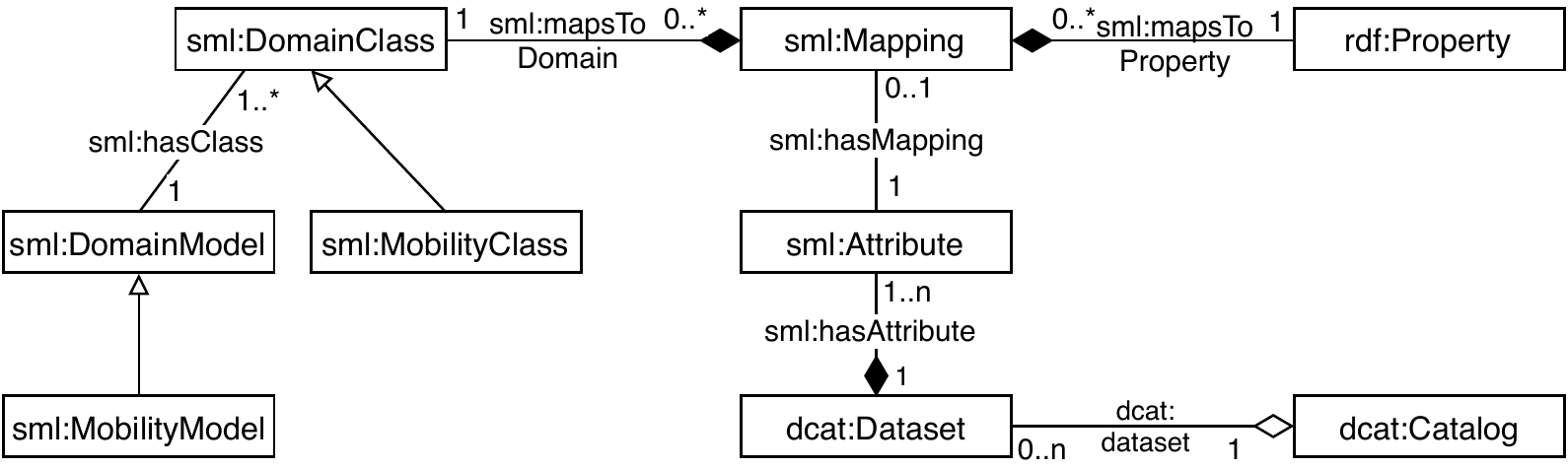}
 \caption{An UML class diagram illustrating the Simple-ML domain model, the data catalog and a partial instantiation of the domain model in the mobility domain.}
 \label{fig:overview_diagram}
\end{figure}

\section{Semantic Models for SDAWs}
\label{sec:approach}

The goals of Simple-ML are realized through a domain model (Fig.~\ref{fig:overview_diagram}), semantic dataset profiles and the SDAW. 
We conduct the modeling in RDF\footnote{Resource Description Framework (RDF): \url{https://www.w3.org/RDF/}} reusing existing vocabularies (e.g. \schema{dcat}\footnote{Data Catalog Vocabulary (DCAT): \url{https://www.w3.org/TR/vocab-dcat/}}), where possible. 
The terms specific to Simple-ML are defined in the Simple-ML vocabulary, denoted using the \schema{sml} prefix.\footnote{The list of the adopted namespaces and the data catalog are available online: \url{https://simple-ml.de/index.php/data-catalog/}}

\textbf{Domain Model:} In Simple-ML, the \textit{domain model} describes relevant concepts, their properties and relations in the specific application domain. 
The class \prefix{sml}{DomainModel} represents the model of an application domain.
The domain-specific concepts are modeled as instances of the class \prefix{sml}{DomainClass}.

\textbf{Dataset Profiles:} \textit{A dataset profile} is a formal representation of dataset characteristics (features). 
\textit{A dataset profile feature} is a dataset characteristic. 
Such features can belong to general, qualitative, provenance, statistical, licensing and dynamics categories \cite{EllefiBBDDST18}. 
In Simple-ML, the goal of the dataset profiles is to define dataset characteristics required to facilitate SDAWs, including information required for data materialization.

\textit{Dataset profile:} A dataset profile is modeled as an instance of \prefix{dcat}{Dataset}. 
General dataset profile features as well as provenance and licensing features
are described using the DCMI Vocabulary (\schema{dcterms}).
Statistical dataset profile features  (e.g. the number of instances) can be  provided at the dataset and the attribute levels.

\textit{Dataset attributes:} 
The attributes of the \prefix{dcat}{Dataset} are modeled as instances of \prefix{sml}{Attribute}. 
An attribute is described through its statistical characteristics at the instance level (e.g. the mean value \prefix{sml}{mean\-Value}), along with the access information to the underlying data source (e.g. the column name in a relational database) to facilitate data access and materialization.

\textit{Dataset access:} 
Simple-ML supports access to datasets 
through dedicated attributes that represent physical storage location and data format (e.g. \prefix{sml}{fileLocation} and \prefix{csvw}{separator}).
Currently, relational databases (\prefix{sml}{Database}) and text files (\prefix{sml}{TextFile}) are supported.

\textbf{Mapping between the Dataset Profile and the Domain Model:} 
Dataset attributes are mapped to the concepts in the domain model (\prefix{sml}{DomainClass}) through the \prefix{sml}{Mapping} class, as illustrated in Fig.~\ref{fig:overview_diagram}. 
This mapping adds domain-specific semantic description 
to the dataset attributes and facilitates their use in the SDAWs.
The class \prefix{sml}{Mapping} provides two properties: \prefix{sml}{mapsTo\-Property} to map a dataset attribute to a property in the domain model, and \prefix{sml}{mapsTo\-Domain} to specify the \prefix{rdfs}{domain} of this property, which is an instance of \prefix{sml}{Domain\-Class}.

\begin{figure}[t]
 \centering
 \includegraphics[width=0.8\textwidth]{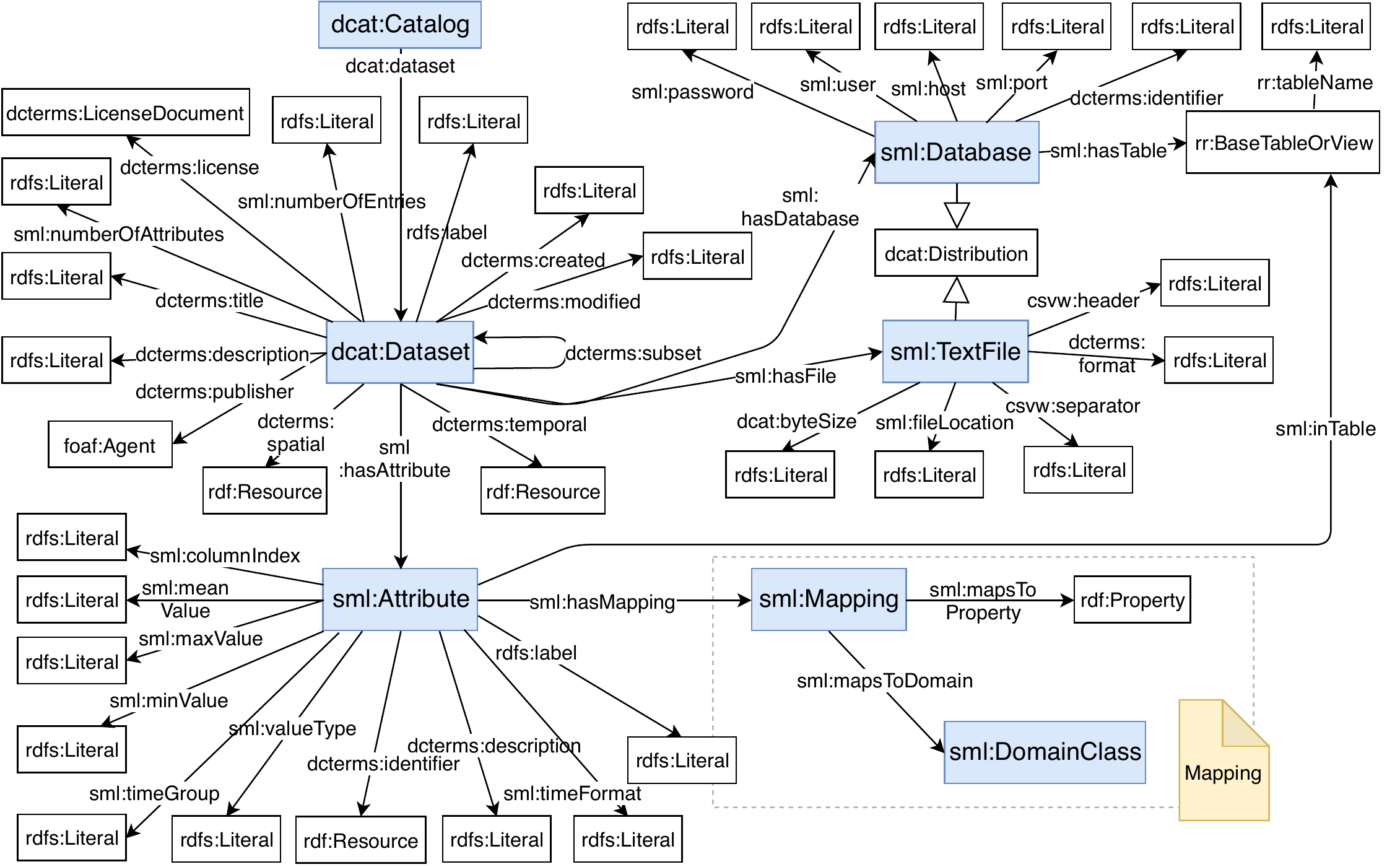}
 \caption{The data catalog schema based on the \schema{dcat} vocabulary. Arrows with an open head denote the \prefix{rdfs}{subClassOf} properties. Regular arrows denote the \prefix{rdfs}{domain} and \prefix{rdfs}{range} restrictions. Blue boxes denote the key \schema{dcat} and \schema{sml} classes.}
 \label{fig:data_catalog_schema}
\end{figure}

\textbf{Data Catalog}: Dataset profiles are organized in a domain-specific data catalog. 
The extensible Simple-ML data catalog is modeled as an instance of \prefix{dcat}{Catalog}. 
The data catalog schema including representations of dataset profiles and the mapping to the domain model is illustrated in Fig. \ref{fig:data_catalog_schema}.

\section{Semantic Data Analytics Workflow (SDAW)}
\label{sec:workflow}

\begin{figure}[ht]
 \centering
 \includegraphics[width=0.8\textwidth]{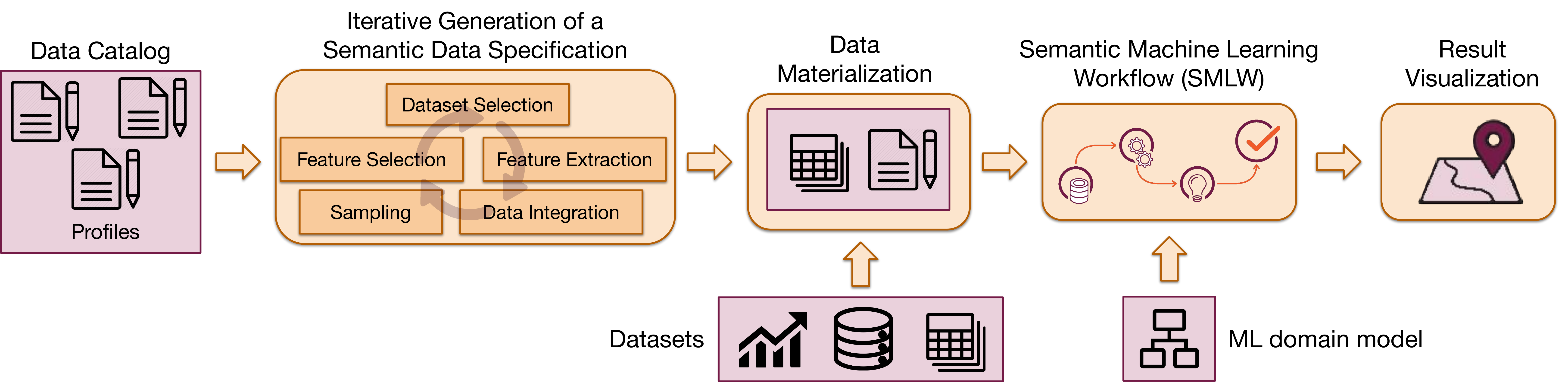}
 \caption{An overview of the Simple-ML Semantic Data Analytics Workflow (SDAW).}
 \label{fig:pipeline2}
\end{figure}

Fig.~\ref{fig:pipeline2} depicts an overview of a \textit{Semantic Data Analytics Workflow} (SDAW). A SDAW consists of several steps discussed in the following.

\textbf{Iterative Generation of a Semantic Data Specification: }
In this first step, the user defines the semantic specification of the data 
to be used in the workflow. 
The input in this step is the data catalog.
The specification is defined through the selection of the operations to be applied to the dataset(s) in the data catalog and their attributes. 
Possible operations include dataset selection, sampling, feature selection, feature extraction and data integration.
These operations can be applied iteratively in a user-defined order.
The Semantic data specification is defined at the metadata level using dataset profiles and does not require any physical data access.
The specification can be stored to facilitate reusability.

\textbf{Data Materialization}: The data specification configured during the previous steps 
is applied to the physical datasets to materialize the integrated data.

\textbf{Semantic Machine Learning Workflow (SMLW)}:
The domain model is complemented with a ML domain model that captures the essential properties of ML concepts and their implementation in specific frameworks. 
A domain specific language (DSL) for SDAWs and SMLWs will include an advanced type system that will use metadata from the application domain to describe datasets and the intermediate results of data processing on one hand, and the metadata of the ML domain to describe the ML processing steps. This will enable statically checking the correctness of applying particular ML methods to particular data. To this extent, we will build upon previous approaches aiming to integrate ontologies into existing type systems (see e.g. \cite{HartenfelsLLS17}). We will go one step further, by designing a language dedicated to the data analytics and ML domain and including data models both for the data and also for the ML processes. 

\textbf{Result Visualization:} 
The domain model can be used to automatically suggest suitable visualizations for specific data types.

\section{Domain Model for Mobility}
\label{sec:domain-model-mobility}

Fig.~\ref{fig:mobility_schema} exemplifies an instantiation of the domain model in the mobility domain. This model includes the following classes: 

\begin{figure}[t]
 \centering
 \includegraphics[width=0.9\textwidth]{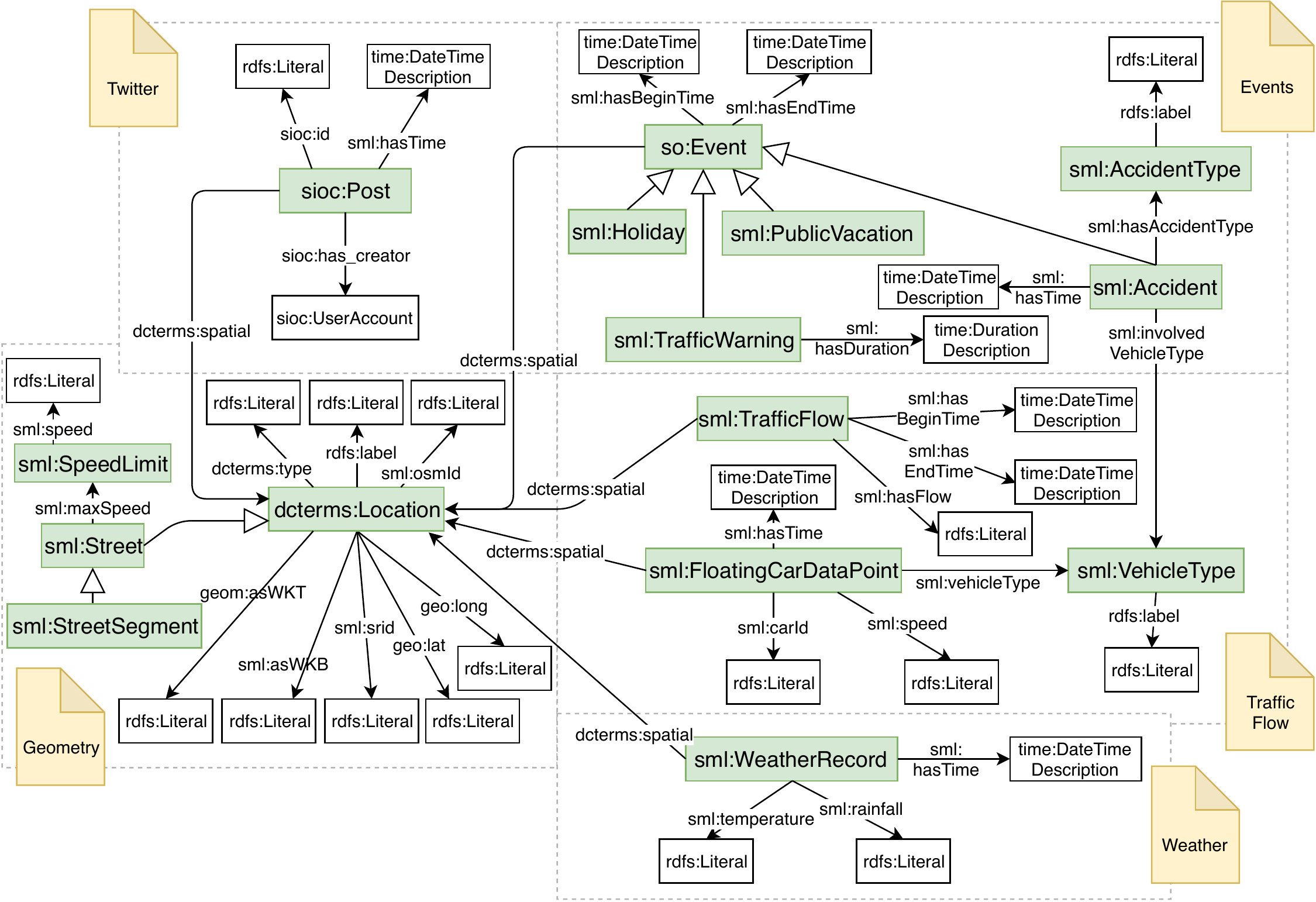}
 \caption{An example domain model for the mobility domain. The arrows with an open head denote the \prefix{rdfs}{subClassOf} properties. Regular arrows denote the \prefix{rdfs}{domain} and \prefix{rdfs}{range} restrictions. Classes in green boxes are sub classes of \prefix{sml}{MobilityClass}.}
 \label{fig:mobility_schema}
\end{figure}

\begin{itemize}
    \item \prefix{sml}{FloatingCarDataPoint}: A vehicle's type, position, time and speed.
    \item \prefix{sml}{TrafficFlow}: Vehicle count statistics (e.g. from 
    road sensors \cite{DBLP:journals/tbd/Nguyen0C17}).
    \item \prefix{so}{Event}: Mobility-relevant events, their time and geographical location.
    
    \item \prefix{sioc}{Post}: Social media posts modeled using the SIOC ontology\footnote{\url{https://www.w3.org/Submission/sioc-spec/}}.
    
    \item \prefix{sml}{WeatherRecord}: Temperature and rainfall 
    at location and time.
    
    \item  \prefix{dcterms}{Location}: Spatial information with geographical coordinates.
    
    \item \prefix{sml}{SpeedLimit}, \prefix{sml}{AccidentType}, \prefix{sml}{VehicleType}: Classes that represent categorical values for speed limits, accident types and vehicle types.
\end{itemize}

These classes are sub classes of \prefix{sml}{Mobility\-Class}, which is a sub class of \prefix{sml}{Domain\-Class} and thus allows the use of \prefix{sml}{Mapping} as shown in Fig.~\ref{fig:data_catalog_schema}.

Fig. \ref{fig:catalog_example} provides an excerpt of an example Simple-ML mobility data catalog.

\begin{figure}[t]
\begin{lstlisting}[showstringspaces=false,basicstyle=\scriptsize\ttfamily,language=sql,deletekeywords={count,YEAR,INTEGER},keywordstyle=\bfseries,frame=single,sensitive=t,morekeywords={FILTER,COUNT}]
sml:SimpleMLCatalog a dcat:Catalog ;
    dcat:dataset sml:FCDDataset .
sml:FCDDataset a dcat:Dataset ;
    dcterms:title "Floating Car Data" ; sml:hasFile sml:FCDDatasetFile ;
    dcterms:temporal [  so:startDate "2017-08-01"^^xsd:date ;
                        so:endDate "2017-12-31"^^xsd:date ] ;
    sml:hasAttribute sml:FCDDatasetAttribute1 .
sml:FCDDatasetFile a sml:TextFile ;
    dcterms:format "text/comma-separated-values" ; csvw:separator ";" .
sml:FCDDatasetAttribute1 a sml:Attribute ;
    rdfs:label "vehicle id"@en ; sml:columnNumber "0"^^xsd:integer ;
    sml:hasMapping [    sml:mapsToProperty sml:carId ;
                        sml:mapsToDomain sml:FloatingCarDataPoint ] .
\end{lstlisting}
\caption{An excerpt of an example data catalog in the mobility domain.}
\label{fig:catalog_example}
\end{figure}

\begin{figure}[t]
\begin{lstlisting}[basicstyle=\scriptsize\ttfamily,language=sql,deletekeywords={count,YEAR,INTEGER},keywordstyle=\bfseries,frame=single,sensitive=t,morekeywords={FILTER,COUNT,PREFIX,OPTIONAL}]
SELECT ?columnNumber  ?attrName ?mapProperty ?mapDomain WHERE {
    sml:FCDDataset sml:hasAttribute ?attribute .
    ?attribute dcterms:identifier ?attrName .
    ?attribute sml:columnNumber ?columnNumber .
    OPTIONAL { ?attribute sml:hasMapping [
        sml:mapsToProperty ?mapProperty ; sml:mapsToDomain ?mapDomain ; ] . } }
\end{lstlisting}
\caption{SPARQL query to select attributes of a given dataset (here: \prefix{sml}{FCDDataset}).} \label{fig:sparql_query_02}
\end{figure}

\section{Simple-ML Application to Traffic Speed Prediction}
\label{sec:application}

We illustrate the iterative generation of a semantic data specification for the problem of traffic speed prediction for a specific road segment at a given time.

\textbf{Dataset Selection:} The user selects a Floating Car Data ($F$) and OpenStreetMap ($O$) datasets. Fig.~\ref{fig:sparql_query_02} shows the SPARQL query to retrieve $F$'s profile.

\textbf{Data Specification}: (i) \textit{Feature Selection}: The user selects four features based on the domain model: \prefix{sml}{maxSpeed}, \prefix{sml}{hasTime} from ($F$) (class \prefix{sml}{FloatingCarDataPoint}), and \prefix{rdf}{type} and \prefix{sml}{maxSpeed} from ($O$) (class \prefix{sml}{StreetSegment}). (ii) \textit{Feature Extraction}: The user selects the following temporal features that are suggested by the system: week day, hour of day from ($F$). (iii) \textit{Data Integration}: A mapping between the vehicle positions in ($F$) and the street segment coordinates in ($O$) is suggested by the system and chosen by the user.

\textbf{Data Materialization}: Using the data specification, relevant features are materialized, with example instances shown in Table \ref{tab:csv_file}. The resulting data can then be used in the SMLW to train a supervised traffic speed prediction model.

\begin{table}[ht]
\small
\caption{Example instances generated using the semantic data specification}
\setlength{\tabcolsep}{0.5em} 
\centering
\begin{tabular}{cclc|lc}
\multicolumn{4}{c|}{\textbf{FloatingCarDataPoint ($F$)}} & \multicolumn{2}{c}{\textbf{StreetSegment ($O$)}} \\ \hline
\textbf{type} & \textbf{speed} & \textbf{time (day)} & \textbf{time (hour)} & \textbf{type} & \textbf{maxSpeed} \ \\ \hline
1 & 74 & Sunday & 23 & motorway\_link & 80 \\
0 & 84 & Sunday & 16 & motorway & \textit{none} \\
1 & 17 & February & 8 & secondary & 70 \\
\hline
\end{tabular}
\label{tab:csv_file}
\end{table}

\section{Related Work}
\label{sec:related_work}

Recent works~\cite{Merkle:2018,EstevesMNSUAL15,HartenfelsLLS17} aim to combine semantics and ML to address a variety of real-world problems. Simple-ML goes one step further and makes use of semantics in the entire DAW. Simple-ML employs dataset profiles and domain-specific data models. The survey~\cite{EllefiBBDDST18} provides a comprehensive overview of RDF dataset profiling methods, tools, vocabularies and features partially utilized by Simple-ML. 
We illustrate the use of Simple-ML in the mobility domain. Mobility has seen many challenges and use cases for data analytics \cite{ijcai2018-482, DBLP:journals/tbd/Nguyen0C17, Tempelmeier-geoinf}. In Simple-ML, the mobility domain is modeled in a light-weight, data-driven manner that facilitates compatibility and reusability of the SDAWs across use cases and datasets.

\section{Conclusion}
\label{sec:conclusion}

In this paper we presented our current development towards the Simple-ML framework. Simple-ML adopts semantic technologies to support the efficient creation, configuration and reusability of robust data analytics workflows. 
We illustrated an application of the framework to a real-world use case in the mobility domain. 
%

\section*{Acknowledgements}
\footnotesize{
This work was partially funded by the Federal Ministry of Education and Research (BMBF), Germany under Simple-ML (01IS18054) and Data4UrbanMobility (02K15A040).

\bibliographystyle{abbrv}
\bibliography{bibliography}
\theendnotes
}

\end{document}